\documentclass[10pt,twocolumn,letterpaper]{article}

\usepackage{cvpr}
\usepackage{times}
\usepackage{epsfig}
\usepackage{graphicx}
\usepackage{amsmath}
\usepackage{amssymb}

\usepackage[pagebackref=true,breaklinks=true,colorlinks,bookmarks=false]{hyperref}

\def\cvprPaperID{3841} 

\cvprfinalcopy

\makeatletter
\def\thickhline{%
  \noalign{\ifnum0=`}\fi\hrule \@height \thickarrayrulewidth \futurelet
   \reserved@a\@xthickhline}
\def\@xthickhline{\ifx\reserved@a\thickhline
               \vskip\doublerulesep
               \vskip-\thickarrayrulewidth
             \fi
      \ifnum0=`{\fi}}
\makeatother

\newlength{\thickarrayrulewidth}
\begin{document}



\title{3D Graph Anatomy Geometry-Integrated Network for Pancreatic Mass\\ Segmentation, Diagnosis, and Quantitative Patient Management}

\author{Tianyi Zhao\thanks{equal contribution}~$^{1,2}$, Kai Cao$^{*3}$, Jiawen Yao$^{1}$, Isabella Nogues$^{4}$, Le Lu$^{1}$, Lingyun Huang$^{5}$, Jing Xiao$^{5}$, \\ Zhaozheng Yin$^{2}$, Ling Zhang$^{1}$\\
$^{1}$PAII Inc.,
$^{2}$Stony Brook University, 
$^{3}$Changhai Hospital,
$^{4}$Harvard University,
$^{5}$Ping An Technology\\
}

\maketitle

\begin{abstract}
\vspace{-2mm}
The pancreatic disease taxonomy includes ten types of masses (tumors or cysts) 
\cite{springer2019multimodality,haj2020pitfalls}. 
Previous work focuses on developing segmentation or classification methods only for certain mass types. 
Differential diagnosis of all mass types is clinically highly desirable \cite{springer2019multimodality} but has not been investigated using an automated image understanding approach.

We exploit the feasibility to distinguish pancreatic ductal adenocarcinoma (PDAC) from the nine other nonPDAC masses using multi-phase CT imaging. Both image appearance and the 3D organ-mass geometry relationship are critical. 
We propose a holistic segmentation-mesh-classification network (SMCN) to provide patient-level diagnosis, by fully utilizing the geometry and location information, which is accomplished by combining the anatomical structure and the semantic detection-by-segmentation network. SMCN learns the pancreas and mass segmentation task and builds an anatomical correspondence-aware organ mesh model by progressively deforming a pancreas prototype on the raw segmentation mask (i.e., mask-to-mesh). A new graph-based residual convolutional network (Graph-ResNet), whose nodes fuse the information of the mesh model and feature vectors extracted from the segmentation network, is developed to produce the patient-level differential classification results. 
Extensive experiments on 661 patients' CT scans (five phases per patient) show that SMCN can improve the mass segmentation and detection accuracy compared to the strong baseline method nnUNet (e.g., for nonPDAC, Dice: 0.611 vs. 0.478; detection rate: 89\% vs. 70\%), achieve similar sensitivity and specificity in differentiating PDAC and nonPDAC as expert radiologists (i.e., 94\% and 90\%), and obtain results comparable to a multimodality test \cite{springer2019multimodality} that combines clinical, imaging, and molecular testing for clinical management of patients. 

\vspace{-4mm}
\end{abstract}


\begin{figure}[t]
\centering
\includegraphics[width=0.9\linewidth]{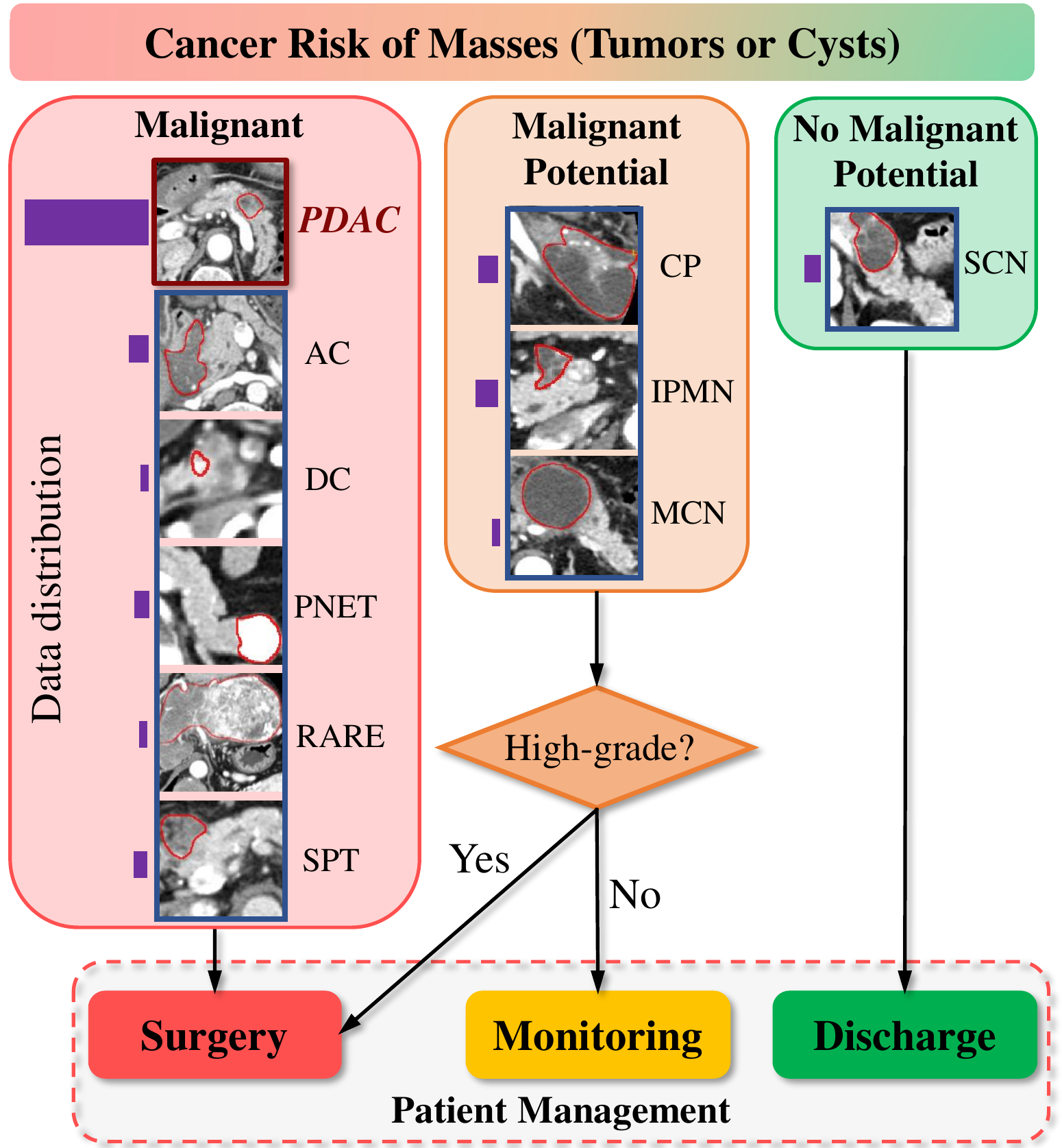}
\caption{Disease taxonomy of the ten types of pancreatic masses (tumors, cysts). Mass type diagnosis determines the clinical malignancy indications that lead to the proper patient risk stratification and management. The purple histogram bars represent their relative frequencies. All images are in the arterial-late CT phase.}
\label{fig:ont}
\end{figure}

\vspace{-1mm}
\section{Introduction}
Pancreatic cancer is the third leading cause of cancer-related deaths in the United States \cite{grossberg2020multidisciplinary}. 
Furthermore, it has the poorest prognosis among all solid malignancies, with a 5-year survival rate $\sim 10\%$ \cite{grossberg2020multidisciplinary,chu2017diagnosis}. Early diagnosis is crucial, as it can potentially increase the five-year survival rate to $\sim 50\%$ \cite{conroy2018folfirinox}. The clinical management of patients with pancreatic disease is based on the potential of the mass to become an invasive cancer. Unlike masses in other organs, pancreatic masses often cannot be reached precisely via needle biopsy due to the pancreas's deep location in the abdomen and the complex network of surrounding organs and vessels. Therefore, reliable imaging-based diagnosis is critical to identifying patients who truly require cancer treatment (e.g., surgery) in a timely fashion, while avoiding unnecessary iatrogenic morbidity. Developing deep learning methods to detect masses, identify malignancies, provide diagnoses and predict cancer prognosis has the potential to revolutionize pancreatic cancer imaging \cite{chu2020pancreatic,chu2017diagnosis,haj2020pitfalls,zhu2019multi,xia2020detecting,zhang2020robust,yao2020deepprognosis}. 

Multi-phase computed tomography (CT) is the first-line imaging modality for the diagnosis of pancreatic diseases. Differential diagnosis of pancreatic masses is challenging for several reasons. (1) 
The same type of mass may appear in different textures, shapes, contrasts, and different enhancement patterns across CT phases. (2) 
Pancreatic ductal adenocarcinoma (PDAC) accounts for most cases in pancreatic cancer specialized hospitals, causing a long-tail problem. (3) 
Masses, at times, are surrounded by inflamed tissues and thus cannot be easily identified. 
The pancreatic diseases in our database encompass ten types of masses (Fig. \ref{fig:ont}): PDAC, ampullary cancer (AC), bile duct cancer (DC), pancreatic neuroendocrine tumor (PNET), rare neoplasm (RARE), solid pseudopapillary tumor (SPT), chronic pancreatitis (CP), intraductal papillary mucinous neoplasm (IPMN), mucinous cystic neoplasm (MCN), and serous cystic neoplasm (SCN).
To this end, we develop anatomy-aware 3D deep graph networks to automatically segment, detect, and perform differential diagnosis of the underlying diseases. 



We tackle two main problems with strong clinical indications: 1) PDAC versus nonPDAC differentiation and 2) clinical management of patients. PDAC is a unique group with the most dismal prognosis. Distinguishing PDAC from nonPDACs is always the primary question to answer. Patient management includes three recommendations: surgery, monitoring, and discharge (lower pannel in Fig. \ref{fig:ont}) \cite{springer2019multimodality}. Patients with malignant masses 
require cancer treatment (e.g., surgery). Those with potentially malignant masses 
require surgery if they are invasive or high-grade dysplasias, or monitoring otherwise. Those with nonmalignant masses could be safely discharged. 
Fine-grained classification of ten classes of masses using multi-phase CT is a very difficult long-tail problem. 


Existing automatic pancreatic mass image analysis methods ~\cite{zhou2017deep,zhu2019multi,zhou2019hyper,zhu2020segmentation,xia2020detecting,zhang2020robust} focus on segmentation of certain types of tumors or cysts and thus cannot exploit the full-spectrum taxonomy of pancreatic mass/disease diagnoses. For pancreatic disease diagnosis, both texture and geometry cues are clinically useful. For instance, some types of masses appear at specific locations of the pancreas: AC and DC appear only at the pancreas head while MCN rarely appears at the head. Others spread over the entire pancreas, such as CP and IPMN. Additionally, some secondary signs of diseases are informative for diagnosis. Parenchymal atrophy and pseudocyst are observed in CP, causing a significant change in the shape of the pancreas. Most pancreatic cancers lead to dilatation of the pancreatic duct, with IPMN in particular abruptly modifying its caliber.

To integrate such prior knowledge/correspondence into the model, (1) we propose a {\bf segmentation based detection network} which can segment and identify pancreatic disease regions simultaneously. Our segmentation network takes multi-phase CT scans as input and outputs segmentation masks of the pancreas (as the studied organ) and mass. We also develop a weak-supervised segmentation method for cases when we have all pixel-level PDAC annotations but only nonPDAC labels. 
(2) A {\bf mask-to-mesh} algorithm is used to build a 3D correspondence-aware mesh from the pancreas segmentation output. The geometry of the pancreas, as well as the location, shape, and distributions of the detected mass can all be captured/encoded by the mesh model. We present a mesh-based feature pooling method that extracts features from the segmentation network and preserve the anatomic structure (each vertex of the mesh has its anatomic meaning). Based on a fixed vertex index list, the pancreas can be automatically divided or parsed into four major sections: head, ventral body, dorsal body, and tail. 
(3) A {\bf geometry-integrated graph classification network} utilizes the 3D anatomy correspondence-aware mesh based deep feature pooling to predict the pancreatic mass type. The network consists of graph-based residual convolutional blocks and an anatomy-based graph pooling layer. All the networks can be trained end-to-end via gradient-based optimization based on a loss function combining the {\bf segmentation loss, mesh vertex classification loss, and global graph classification loss}.  

Our main contributions are three-fold. (1) To the best of our knowledge, this is the first work to propose a multi-phase CT imaging analysis method for the full-spectrum taxonomy of pancreatic mass/disease diagnosis. (2) We are the first to integrate the 3D geometry-aware mesh model for effective pancreatic mass (tumor or cyst) imaging analysis (Sec.~\ref{sec:anatomy}, Sec.~\ref{sec:Graph-ResNetwork}), explicitly capturing the anatomy-mass integrated geometry and texture cues. 
(3) We have extensively evaluated our models on 661 patients (five-phase CT scans per patient). We achieve a new state-of-the-art PDAC segmentation accuracy (Dice: 0.738) and a substantially better nonPDAC segmentation (Dice: 0.611 vs. 0.478) and detection accuracy (detection rate: 89\% vs. 70\%) compared to the strong baseline method nnUNet. Our imaging-only automated approach 
demonstrates comparable performance levels with (a) expert radiologists in the differentiation of PDAC versus nonPDAC who combine the analysis on clinical factors, imaging, and blood tests, and (b) a state-of-the-art machine learning-based clinical patient management system (i.e., surgery, monitoring, discharge) using the multimodality tests \cite{springer2019multimodality}. 




\begin{figure*}
\centering
\includegraphics[width=0.92\linewidth]{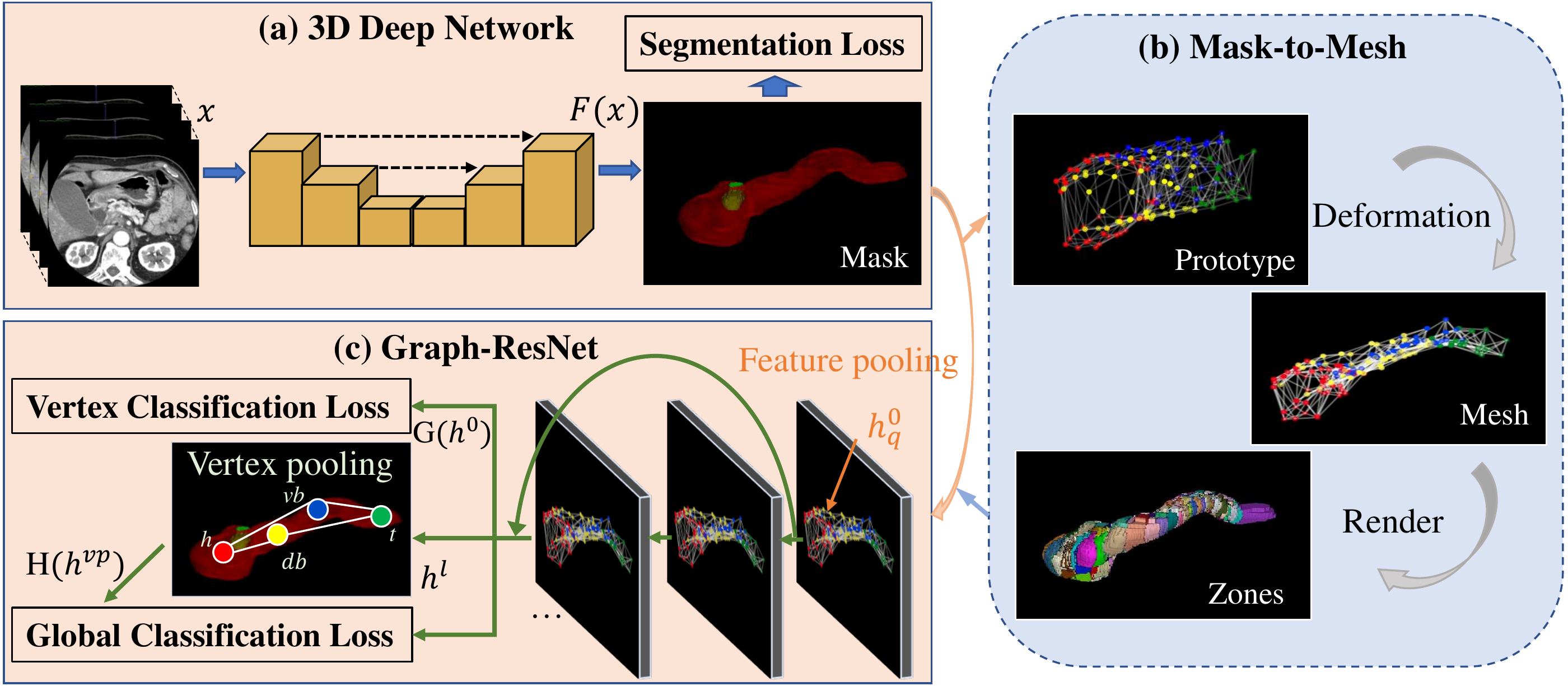}
\caption{Flowchart of our proposed Segmentation-Mesh-Classification Network (SMCN). SMCN has three components: the pancreas-mass segmentation network, the mask-to-mesh 3D anatomy modeling, and the global mass classification network. The mesh model is the bridge between the segmentation network and the classification network, who pools the features from the segmentation network into the vertex feature vectors in the graph classification network.}
\label{fig:e2e} \vspace{-3mm}
\end{figure*}

\section{Related Work}


{\bf Deep Mesh Learning.} Deep 3D reconstruction has been extensively studied in the computer vision and graphics fields~\cite{chang2015shapenet, huang2015single,Wang_2018_ECCV,Wen_2019_ICCV}, and various methods have been proposed to learn 3D shapes of organs from medical images~\cite{ni2019elastic,yao2019integrating,Wickramasinghe2020Voxel2Mesh3M}. 
The key component of mesh learning methods is the graph convolutional neural network (GCN)~\cite{Kipf:2016tc}, typically used for graph-structured data processing.
A liver mesh modeling method, inspired by the Pixel2Mesh algorithm \cite{Wang_2018_ECCV}, is proposed in \cite{yao2019integrating}, which simultaneously generates the mesh model and segmentation mask with improved geometry and segmentation accuracies. Voxel2Mesh \cite{Wickramasinghe2020Voxel2Mesh3M} learns the mesh of the liver directly from 3D image volumes with a new mesh unpooling operation for better mesh reconstruction. However, all of these methods are designed specifically for organ segmentation, and it is not clear how their segmentation and mesh results could be used for diagnosis. 

{\bf Pancreatic Image Analysis.} When manual segmentations of masses are available, radiomics schemes are commonly used to classify disease types \cite{chu2019application,dalal2020radiomics,chu2020pancreatic}. However, heavy reliance on hand-annotated masks can make radiomics models less reproducible and scalable. To achieve automation, researchers have used detection-by-segmentation networks \cite{zhu2020segmentation,zhu2019multi,xia2020detecting,zhou2019hyper} with U-Net as a common backbone network  \cite{ronneberger2015u}. More recently, nnU-Net \cite{nnunet} (a self-adapting framework based on vanilla U-Nets) and its self-learning version have achieved competitive accuracy on PDAC segmentation \cite{DBLP:journals/corr/abs-1902-09063,wang2020deep,zhang2020robust}.  Shape-induced information, e.g., tubular structure of dilated duct, has been exploited along with the PDAC segmentation task \cite{liu2019joint,xia2020detecting,wang2020deep} to improve diagnosis.
Existing studies mainly focus on PDAC \cite{zhu2019multi,xia2020detecting,xia2020detecting,zhou2019hyper, chu2019application, weisberg2019deep} or PNET \cite{zhu2020segmentation}, which cannot fully meet routine clinical needs on the full taxonomy of pancreatic tumor diagnosis. 
A comprehensive taxonomy for pancreatic diseases has been clinically defined to help managing/treating patients~\cite{springer2019multimodality} where a machine learning model is built on top of clinical factors, imaging characteristics, and molecular biomarkers. These complex measurements require intensive labor costs and manual intervention, with reduced generalization and reproducibility.




\begin{figure*}
\centering
\includegraphics[width=0.9\linewidth]{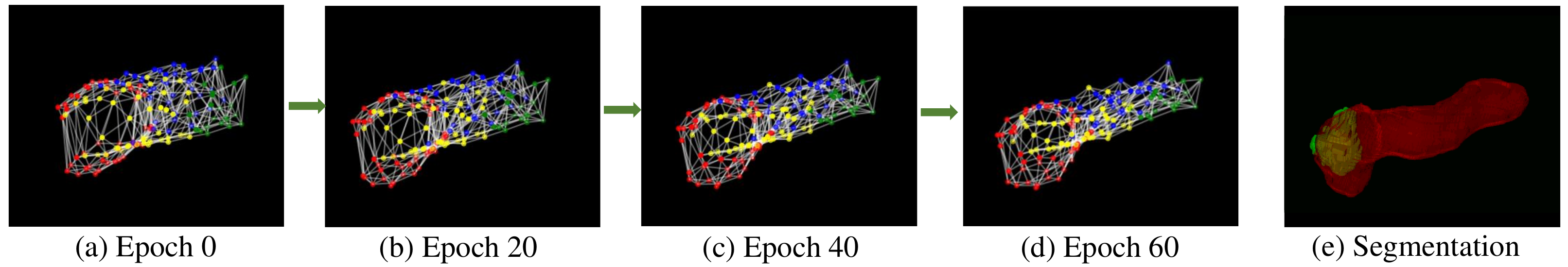}
\caption{The 3D mesh deformation process. The mesh is initialized as the pancreas prototype (a). Given the pancreas-mass segmentation mask (e), the geometry of the mesh is deformed gradually to fit the surface of the segmentation mask, as in (b-d). Red: the head of  pancreas; Blue: the ventral body of pancreas; Yellow: the dorsal body of pancreas; Green: the tail of pancreas.}
\label{fig:meshfitting} \vspace{-3mm}
\end{figure*} 

\section{Methods}

Our segmentation-mesh-classification framework 
is illustrated in Fig. \ref{fig:e2e}, with three main components as below.

\subsection{Anatomy-Mass Segmentation Network}
\label{sec:seg}
The inputs $X$ of the segmentation network are multi(5)-phase 3D CT scans, 
which are concatenated into a 4D input, $X \in R^{5\times W\times H\times D}$. The input $Y$ is the label/annotation including the auto-segmented pancreas (by an ensemble nnUNet \cite{nnunet} model trained on a public pancreas dataset \cite{DBLP:journals/corr/abs-1902-09063}) and radiologist-segmented PDAC and nonPDAC masses. It is represented by the one hot encoding $Y \in R^{K\times W\times H\times D}$. The numbers of labels $K$ can differ by task, i.e. PDAC vs. nonPDAC or patient management. Our backbone network is nnUNet.The pixel-level segmentation loss combines the cross-entropy and Dice losses: 
\vspace{-3mm}
\begin{equation}
\label{eq:pixel}
\begin{split}
    L_{CE} &= - \sum_{w,h,d} \sum_k^K y_{k,w,h,d}\; log(F(x)_{k,w,h,d}),\\
    L_{DC} &= -2 \sum_k^K \frac{\sum_{w,h,d}F(x)_{k,w,h,d} y_{k,w,h,d}}{\sum_{w,h,d}F(x)_{k,w,h,d} + \sum_{w,h,d}y_{k,w,h,d}},\\
    L_{Seg} &= L_{CE} + L_{DC}.
\end{split}
\end{equation} 
\noindent where $F(x)$ is the softmax output of the network, $F(x) \in R^{K\times W\times H\times D}$. 

NonPDAC labels are usually more difficult to obtain than PDAC due to their diversified appearance in a long-tail distribution. As such, we can perform the PDAC vs. nonPDAC task using only the PDAC mask annotations (when nonPDAC ones unavailable) as follows. First, we train a segmentation model on the PDAC data alone, focusing on PDAC segmentation. Next, we apply the final PDAC segmentation model to the nonPDAC data. A large portion of the nonPDAC masses are solid tumors that can possibly be misdetected as PDAC. All raw detections from the nonPDAC dataset by the PDAC segmentation model are collected as pseudo nonPDAC labels. Finally, we train a new segmentation network by using both original PDAC and pseudo nonPDAC labels.

\subsection{3D Mesh-based Anatomy Representation} \label{sec:anatomy}

Unlike existing mesh learning methods~\cite{Wang_2018_ECCV,Wickramasinghe2020Voxel2Mesh3M} that are initialized by a randomized ellipsoid mesh, we encode the prior knowledge into our initial anatomy mesh by fitting it to the pancreas shape with anatomical meanings.

{\bf Pancreas Anatomy.} We first create a prototype mesh based on the average pancreas shape from the training fold. Afterwards, we place 156 vertices that are equally distributed on the surface of the prototype to build an anatomic structure. In particular, we separate them among the four anatomical regions of the pancreas, namely the pancreas head, ventral body, dorsal body and pancreas tail. The first 48 vertices belong to the pancreas head, denoted as red dots in Fig. \ref{fig:meshfitting}. The 49-90th vertices belong to the ventral body of the pancreas, depicted by blue dots. The 91-135th vertices correspond to the dorsal body of the pancreas, shown in yellow dots. The last 21 vertices compose the tail of the pancreas, illustrated by green dots. 

{\bf Mask-to-mesh Process.} 
The next step is to deform the mesh to pancreas mask of each patient as the target (Fig. \ref{fig:meshfitting}-(e)). The geometry of the mesh can be deformed gradually to fit the surface of the segmentation mask, as in Fig. \ref{fig:meshfitting}(a-d), so as to preserve the true anatomy of the pancreas.
In this way, our mesh could maintain the anatomical meaning after deformation. 
To guide this deformation process, we define a loss function composed of three terms: point loss and two edge regularization terms. We define $p$ to be the vertex in the mesh and $q$ the voxels of the surface in the segmentation mask. {\bf Point loss}, intuitively,  measures the distance of each point to the nearest point at the surface of the segmentation. $L_{pt} = \sum_p min_q || p-q||_2^2.$
With the point loss, the pancreas mesh can be driven to fit the segmentation mask. 
Then we propose the  {\bf first edge regularization term} in order to preserve the geometry of the mesh:
$L_{e_1} = \sum_e  || e-mean(e)||_2^2, \; e= || p-p'||_2, \; p' \in N(p)$
where $N (p)$ is the neighboring vertices of $p$.
Next, to penalize the flying vertices (i.e. abnormal vertices randomly updated during the deformation process, resulting in structure flaws), we propose the {\bf second edge regularization loss} to simply minimize the edge length: $L_{e2} = \sum_e e$. 
Finally, the overall loss is \vspace{-2mm}
\begin{equation}
L_{meshfit} = L_{pt} + \lambda_1 L_{e1} + \lambda_2 L_{e2}
\end{equation}
Note that mesh vertices can keep their anatomic meaning even after the deformation process. The mesh fitting process can automatically parse the head, ventral body, dorsal body, and tail of the pancreas and could better preserve the anatomical-aware information in the mesh deformation. 

{\bf Rendering.} Based on the coordinates of the mesh vertices, we divide the pancreas into zones $Z(p)$. Each voxel of the segmented pancreas volume is defined in a zone by its nearest vertex, as shown in Fig. \ref{fig:e2e}(b). The rendering method to define the zones from the each vertex includes the following steps: 1) The vertex in a 3D volume $A \in R^{W,H,D}$ is labeled by its index, i.e., the $i$th vertex at $(w,h,d)$ is labeled as $i$, $A_{w,h,d} = i$. All the other voxels of $A$ are set to zero. 2) The 3D volume $A$ is dilated by one voxel as $A'$. Each zero-voxel in $A$ with a corresponding non-zero voxel in $A'$ that belongs to the pancreas in the segmentation output $F(x)$ is substituted with the corresponding voxel in $A'$.  3) Repeat the step 2) until all voxels have been updated.





\subsection{Global Mass Classification Network} \label{sec:Graph-ResNetwork}

\begin{figure}
\centering
\includegraphics[width=0.9\linewidth]{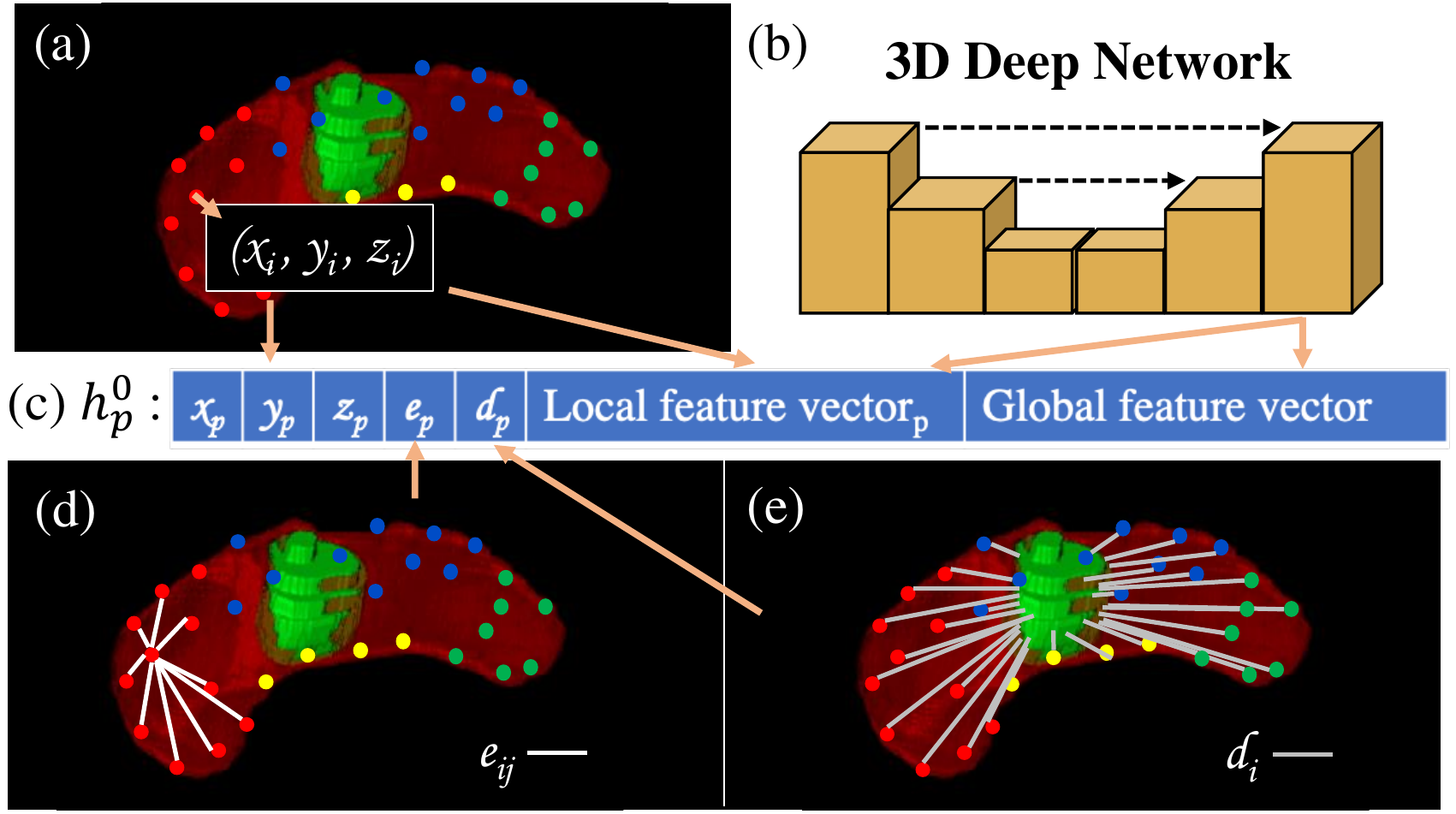}
\caption{Illustration of the feature vector $h_i$ for the  $i$th vertex. $x_i, y_i, z_i$ represents the location of the $i$th vertex. $e_i$ is the average of all edge lengths $e_{ij}$ (i.e., the distance from every neighboring vertex $j$ of to $i$). $d_i$ denotes the shortest distance from the $i$th vertex to the tumor surface. }
\label{fig:h} \vspace{-3mm}
\end{figure}

{\bf Vertex Feature Pooling.} 
Given a deformed mesh and its representative vertex zones, we integrate them into the deep network for shape-constrained detection and segmentation.
We encode the anatomical structure of the pancreas, as well as the tumor's texture and geometry into the feature vector. As such, the feature vector can be viewed as an anatomical representation of the pancreas and tumor of interest.
We define  $h_p^0$ as the initialized feature vector attached to vertex $p$ of the pancreas mesh, as shown in Fig. \ref{fig:h}(a,c). The anatomical feature representation contains the vertex coordinates $(x_p, y_p, z_p)$, the average edge length to its neighbours $e_p$, the distance from $p$ to the nearest tumor surface point $d_p$, and the local and global feature vectors pooled from prediction probability maps of the segmentation network.  
More specifically, $e_p = \sum_{p' \in N(p)} e_{pp'}/| N(p)|$ (Fig. \ref{fig:h}(d)) and 
$d_p = min_r ||p-r||^2_2$, where $r$ is the point at the surface of the mass  (Fig. \ref{fig:h}(e)). 
The zone $Z(p)$ then acts as the receptive field within which we pool the prediction probability maps of the 3D segmentation network ($F(x)$) to obtain a local feature vector, as shown in Fig. \ref{fig:h}(b). The probability maps within the receptive field can be represented as a list of feature vectors $F(x)_{w,h,d} \in R^{K}$, where $(w,h,d) \in Z(p)$.

In addition to Mesh-based Feature Pooling, we also perform global average pooling to get a global feature vector from the probability map of the 3D segmentation Network in Fig. \ref{fig:h} (b). We then concatenate the meta information, the local feature vectors of each individual vertex, and the global feature vector of the entire mesh to create the feature vector $h_p^0$. We explicitly encode the pancreas-to-tumor geometry, the tumor's location in the pancreas, and its texture in the mesh-based anatomy representation. This improves the model's ability to learn anatomy-aware features, which will ultimately improve its performance in pancreatic mass/disease diagnosis based on the full-spectrum pancreatic disease taxonomy.


{\bf Graph-ResNet.} After obtaining feature vectors $h_i$ for each vertex, we feed them into the graph-based residual convolutional network (Graph-ResNet) which has proven to be successful in a recent deep mesh learning application ~\cite{Wang_2018_ECCV}. 
We reduced the original Graph-ResNet to a smaller network containing six graph convolutional layers and shortcut connections between every two layers to perform the tumor classification task under three granularity levels.  
Each graph-based convolutional layer is defined as:
\begin{equation}
    h^{l+1}_p = w_0 h^{l}_p + \sum_{p' \in N(p)} w_1 h^l_{p'}
\end{equation}
where $h^l_p$ is the feature vector attached on vertex $p$ at layer $l$ of the Graph-ResNet. $w_0$ and $w_1$ are learned parameters, with $w_1$ being shared by all edges.
The graph-based convolutional layer accounts for the way in which vertices neighboring a given vertex regularize the vertex-to-neighbor exchange of information.
We use two classification training losses: the vertex-level classification loss and the global classification loss. 
The {\bf vertex classification loss} is defined as the cross-entropy loss and is applied to each vertex as 
\begin{equation}\label{eq:vertex}
    L_{Vertex} =  - \sum_{p} \sum_k^K y^v_{k,p}\; log(G(h^0)_{k,p}),
\end{equation}
where $G(h^0)$ denotes the softmax output of the graph network at every vertex. The vertex label $y^v$ is inferred from the labeled mask. 
Background voxels are labeled as 0,  pancreas voxels as 1, and voxels for all mass types with labels greater than 1. The vertex $q$ is labeled using the maximum value of the voxels in its corresponding zone $Z(p)$, $\hat{y}^v_p = max_{(w,h,d) \in Z(p)} \hat{y}_{w,h,d}$. 
$y^v$ is the one-hot encoding of $\hat{y}^v$.
Considering that some mass types may have blurry boundaries and the mass neighbor $Z(p)$ may also contain relevant cancer-related information, 
this labeling strategy and the propagation/smoothing property of the Graph-ResNet makes our approach robust to the quality of segmentation annotations or labels and increase the detection rates of masses.

{\bf Global Mass Classification.}
After running Graph ResNet, we pool four features from all four pancreatic regions according to the vertex indices (1-48, 49-93, 94-135, 136-156), as shown in Fig. \ref{fig:e2e}. 
These four global feature vectors and 156 local feature vectors are concatenated into one vector $h^{vp}$, and fed into the final mass classification layer. The final global classification layer is a fully-connected layer. The global classification loss is the cross-entropy loss:
\vspace{-2mm}
\begin{equation}
    L_{Global} = -  \sum_k^K y^g_{k}\; log(H(h^{vp})_{k}),
\end{equation}
where $H(h^{vp})$ is the softmax output of the classification layer, and $y^g$ the patient-level mass/disease label. 
The overall loss function is the combination of three losses:  \vspace{-2mm}
\begin{equation}
    L = L_{Seg} + \eta_1 L_{Vertex} +  \eta_2  L_{Global}.
\end{equation}
where $\mathbf{\eta}$ is a hyperparameter used to balance the three loss components. $L_{Seg}$ is the pixel-level segmentation loss 
, and $L_{Vertex}$ is the vertex classification loss (Eq.\ref{eq:vertex}). All networks, namely the  3D segmentation network and the classification Graph-ResNet, can be trained end-to-end by leveraging this global loss function.

\section{Experiments}
{\bf Data and Preprocessing.} Our dataset contains 661 patients with surgical pathology-confirmed pancreatic masses (366 PDACs, 46 ACs, 12 DCs, 35 PNETs, 13 RAREs, 32 SPTs, 43 CPs, 61 IPMNs, 7 MCNs, and 46 SCNs). 
Each patient has 5-phase CT scans: non-contrast (NC), arterial-early (AE), arterial-late (AL), venous (V), and delay (D). The median voxel size is $0.419\times 0.419 \times 3$mm. The manual annotations of masses was performed by an experienced pancreatic imaging radiologist on the AL phase; 
CT scans from the other four phases are registered to AL by DEEDS \cite{Heinrich2013Deeds}. 
Data augmentation is performed on-the-fly. This includes spatial transforms, Gaussian blur, and contrast shifting \cite{nnunet}. The hyper-parameters are 
set as $\lambda_1 = 10^{-4}$, $\lambda_2 = 10^{-2}$, and $\eta_1 = \eta_2 = 0.1$ based on preliminary experiments. 
All experiments are performed using nested three-fold cross-validation. In each fold, the network is trained for 1000 epochs. The best model is selected based on its performance on the validation set. It is then applied to the test set to generate the final experimental results.

\subsection{Evaluation on Mass Segmentation}


{\bf Quantitative Evaluation.} Segmentation accuracy is measured using the Dice coefficient. Dice scores and detection rates of PDACs and nonPDACs (10 disease classes in total) are provided in Table \ref{tab:10dice}. A detection is considered successful if the intersection (between the ground truth and segmentation mask) over the ground truth is $\geq 10\%$ (counted as 1); otherwise, it is considered a misdetection (counted as 0). 
Our SMCN network is compared to a strong baseline, 3D nnUNet \cite{nnunet}, trained from scratch on our dataset. We find that integrating the 3D mesh-based anatomy representation into Graph-ResNet substantially improves mass segmentation and detection accuracies relative to nnUNet's, especially for nonPDAC (Dice: 0.611 vs.  0.478;  Detection rate:  88.7\% vs.  69.3\%).
\begin{figure} [t]
\centering
\includegraphics[width=\linewidth]{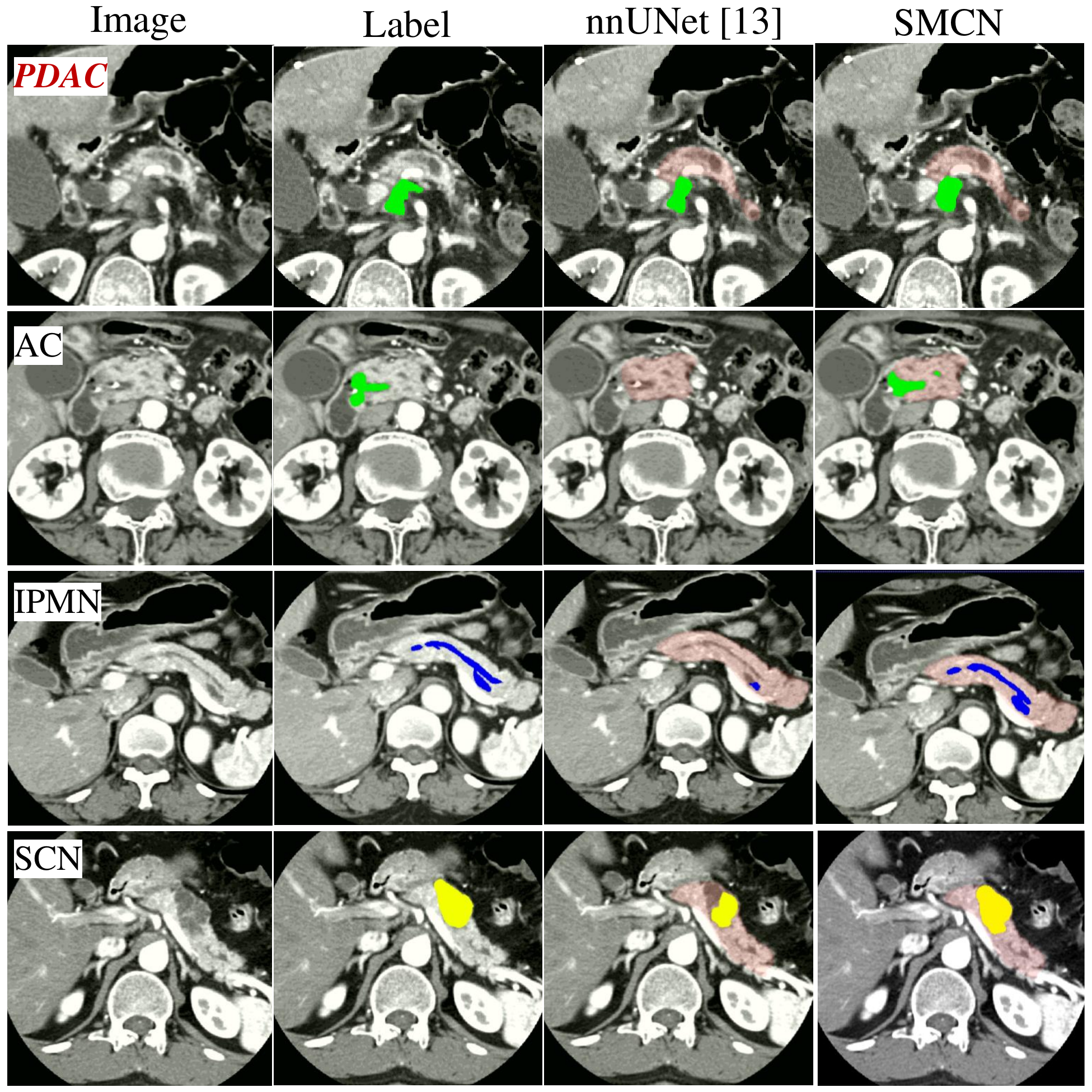}
\caption{Examples of pancreas-mass segmentation results. Light red: pancreas. Green: surgery mass. Blue: monitoring mass. Yellow: discharge mass.}
\label{fig:segs} \vspace{-3mm}
\end{figure}
\begin{figure}
\centering
\includegraphics[width=\linewidth]{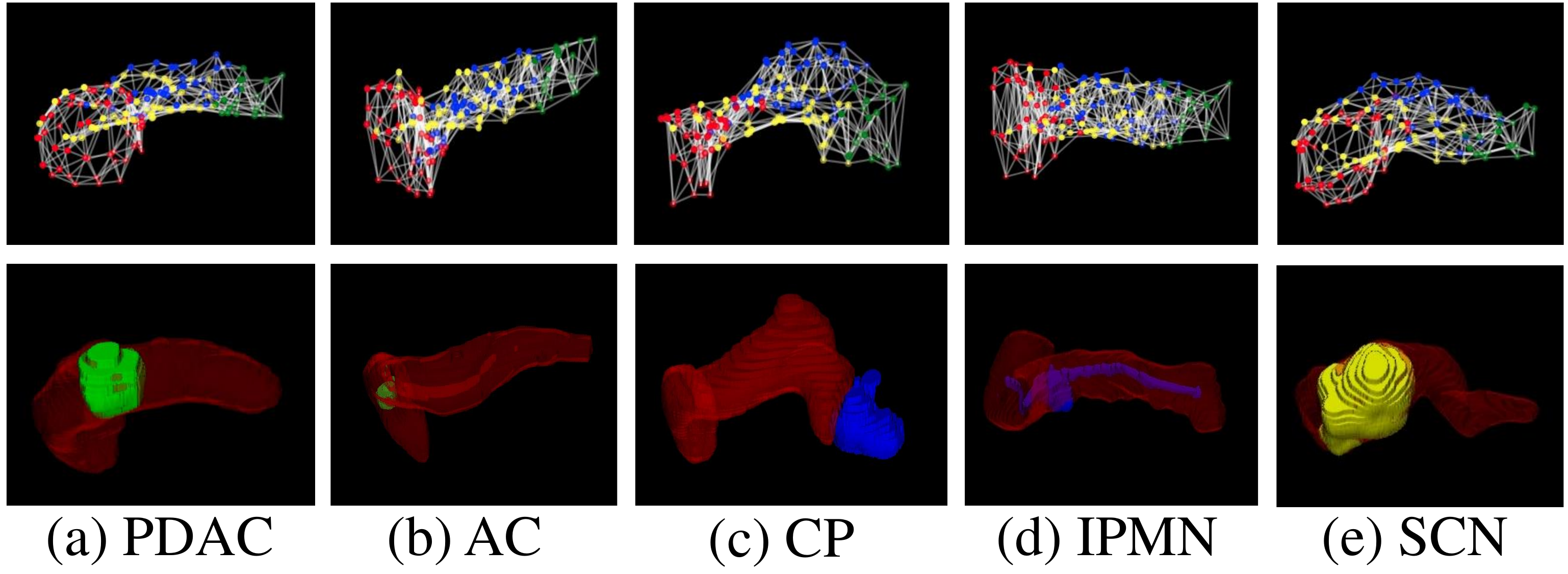}
\caption{Examples of 3D meshes generated from pancreas segmentation under various scenarios. {\bf Top:} red denotes the head of pancreas; blue the ventral body; yellow the dorsal body; and green the tail. {\bf Bottom:} light red depicts the pancreas; green the surgery tumor; blue the monitoring tumor; and yellow the discharge tumor.}
\label{fig:meshs} \vspace{-3mm}
\end{figure}

{\bf Qualitative Evaluation.} Fig.\ref{fig:segs} illustrates the segmentation results for qualitative comparison. SMCN can guide the segmentation network by integrating geometry cues and achieve more complete segmentation results, especially for harder and relatively rare nonPDAC classes. In Fig.\ref{fig:segs}, nnUNet misdetects an AC (malignant) completely because the tumor is outside of the pancreas head. Our SMCN model instead detects this AC finding. For an IPMN case (malignant potential), our model provides more complete segmentation including the clinically-critical secondary signs of pancreatic duct dilatation, than the 3D nnUNet model \cite{nnunet}. For SCN (benign), SMCN segments the full tumor surface. All demonstrate that our anatomy geometry-aware SMCN network produces superior segmentation results against the state-of-the-art 3D nnUNet \cite{nnunet},  being widely used in medical imaging applications.


{\bf PDAC Segmentation under Different CT Phases.} We compare our method to three other state-of-the-art methods of \cite{zhang2020robust}, DDT \cite{wang2020deep}, and \cite{xia2020detecting} which are applied to Venous phase. For ease of comparison, we implement/evaluate our models using only one or a few CT phases (in Table \ref{tab:dice}). 
The AL phase outperforms all other individual phases due to its higher voxel resolution and imaging contrast; the second best is Venous phase, widely adopted in clinical settings. 
 The fusion/concatenation of multiple CT phases as the network's input yields better results than any single phase. The combined five phases yield the best segmentation Dice score of 0.738, which noticeably improves upon the previous state-of-the-art result of 0.709 \cite{zhang2020robust}. Note that in \cite{zhang2020robust} PDACs are mostly small in size and are only located at the pancreas head. In our dataset, by contrast, PDACs appear in various sizes and span the entire pancreas. The fusion of all four phases but AL as input yields better results than any single phase as well (Dice scores 0.696 vs. $0.562\sim0.675$). 

\begin{table*}
\centering
\caption{Segmentation and detection results over the ten types of pancreatic masses. Note that in this experiment, networks are trained with four labels: background, pancreas, PDAC, and nonPDAC. micro: result for all patients; macro: average of the metrics of the ten classes. }
\label{tab:10dice}
\resizebox{\textwidth}{!}{
\begin{tabular}{c| c  | c|  c|c|c|c|c|c|c|c|c|c|c|c }
\thickhline
&Metric     & PDAC & NonPDAC &AC & CP& DC & IPMN & MCN  & PNET & RARE & SCN & SPT  & micro & macro \\
\hline
&Average Dice & 0.734& 0.478 & 0.423& 0.533& 0.001 & 0.151& 0.924 & 0.514& 0.849 & 0.585 & 0.762 & 0.618 &0.548 \\
nnUNet\cite{nnunet}&Detection rate & 0.959& 0.693 & 0.7& 0.778& 0.0 & 0.308 & 1.0& 0.833 & 1.0 & 0.8  & 1.0  & 0.838 & 0.738 \\
\hline
SMCN& Average Dice & 0.738& 0.611& 0.438& 0.668 & 0.006 & 0.602& 0.924 & 0.514 & 0.824 & 0.666 &0.801 & 0.681 &0.618 \\
(Ours)&Detection rate & 0.972& 0.887& 0.8& 0.889& 0.0 & 1.0 & 1.0& 1.0 & 1.0 & 0.8  & 1.0  & 0.934 & 0.846 \\
\thickhline

\end{tabular}
} \vspace{-3mm}
\end{table*}

\begin{table}
\centering
\caption{PDAC segmentation under different CT phases: non-contrust (N), arterial (A), arterial-early (AE), arterial-late (AL), venous (V), and delay (D).}
\label{tab:dice}
\begin{tabular}{ l    c   c}
\hline 
\hline 
 Methods  &  CT Phases  &  Average Dice\\
\hline
\multicolumn{3}{c}{Reported from their original papers}\\
\hline
Zhang et al.   \cite{zhang2020robust}  & N+A+V & 0.709  \\
DDT  \cite{wang2020deep}  & V &  0.634 \\
Xia et al. \cite{xia2020detecting}   & A+V & 0.644  \\
\hline
\hline
Ours  &  AE &0.646    \\
Ours  &  AL &0.730     \\
Ours  &  V  &0.675     \\
Ours  &  D  &0.562      \\
Ours  &  N+AE+V+D  &0.696     \\
Ours  &  N+AE+AL+V+D  &0.738 \\
\hline
\end{tabular}
\end{table}

\subsection{3D Anatomy Mesh Quality Evaluation}

Qualitative results of our generated mesh model are shown in Fig.~\ref{fig:meshs}. Our method can effectively and accurately model the diseased pancreases under various scenarios, e.g., PDAC at the neck (Fig.~\ref{fig:meshs}(a)), AC at the head with pancreatic duct dilatation (Fig.~\ref{fig:meshs}(b)), pseudocyst at the tail (Fig.~\ref{fig:meshs}(c)), IPMN on the full pancreatic duct (Fig.~\ref{fig:meshs}(d)) and large tumor and parenchymal atrophy (Fig.~\ref{fig:meshs}(e)). As described in Sec.~\ref{sec:anatomy}, our 3D mask-to-mesh algorithm can automatically divide the pancreas into four parts: head, ventral body, dorsal body and tail, which are color-coded as red, blue, yellow and green. The distributions of anatomic locations  of 10-class pancreatic masses in our dataset are described in the supplementary material. AC and DC only appear at the head; MCN does not appear on the head; CP and IPMN appear mostly on the head; PDAC, PNET and RARE are distributed over the entire surface of the pancreas. Cystic masses (SPT, SCN and MCN) are distributed over the entire pancreas with the majority being located at the dorsal body. These observed distributions verify the prior knowledge of different masses' spatial locations and motivate our geometry-aware mass diagnosis framework. 

From the generated 3D mesh model, we can accurately measure the shape and location of the mass against the pancreas, as shown in Fig.\ref{fig:h}, at a human-readable level (156 is a suitable number for the clinical expert's perception). Our proposed mask-to-mesh algorithm and SMCN models can be generalized to other important organs with specific anatomical structures (like liver), which can potentially improve the underlying imaging-based mass diagnosis.


\begin{table}
\centering
\caption{Average values of PDAC vs. nonPDAC across 3 folds. PV: pixel voting; VV: vertices voting; GC: global classification.}
\label{tab:acc2}
\begin{tabular}{ c  c  c c  }
\hline 
\hline
Methods   & Accuracy  &Sensitivity &  Specificity   \\
\hline
Radiomics~\cite{van2017computational}  &0.857   &0.853    &   0.868            \\
DeepTexture~\cite{xue2018deep}  &0.770   &0.827    &   0.697          \\
ResNet3D~\cite{hara3dcnns}  &0.720   &0.788    &   0.633         \\



SMCN  w PV  &0.913     & 0.945      &   0.875          \\
SMCN w VV  &  0.920   &0.945    &   0.891              \\
SMCN w GC  &  0.927   &0.945    &   0.906           \\
Expert radiologists & - & $\sim$0.94 & $\sim$0.90 \\
\hline
\end{tabular} \vspace{-3mm}
\end{table}

\begin{table*}[t]
\centering
\caption{Classification confusion matrix of quantitative patient management and comparison our results against the clinical test   \cite{springer2019multimodality}.}
\label{tab:cmfu}
\begin{tabular}{ c|c|c|c|c |c |c |c }
\hline 
CompCyst \cite{springer2019multimodality} &Discharge    & Monitoring  & Surgery & Ours &  Discharge   & Monitoring  & Surgery  \\
\hline




Discharge (n=53) &32 (60\%) &14 (26\%)    & 7 (13\%) & (n=46) &28 (61\%) &10 (22\%)    & 8 (17\%)\\
\hline
Monitor (n=140) &1 (1\%) &68 (49\%)    & 71 (51\%) & (n=104) &11 (11\%) &65 (63\%)    & 27 (26\%)\\
\hline
Surgery (n=152) & 0 (0\%) &14 (9\%)    &138 (91\%) & (n=440) & 6 (1\%) &16 (4\%)    &418 (95\%)\\
\hline
\end{tabular} \vspace{-4mm}
\end{table*}

\begin{table}
\centering
\caption{Experimental Results for Patient Management. PV: pixel voting; VV: vertex-based voting; GC: global classification.}
\label{tab:acc3}
\begin{tabular}{ c    c  c  }
\hline 
\hline
Methods  &   Accuracy   &   Average Recall  \\
\hline

Radiomics~\cite{van2017computational}   & 0.854    &  0.667         \\
DeepTexture~\cite{xue2018deep}  &0.743      &  0.557         \\
ResNet3D~\cite{hara3dcnns}  &0.737      &  0.607         \\



SMCN  w PV  &0.832     & 0.735              \\
SMCN  w VV  &0.839     & 0.656             \\
SMCN w GC       &0.865      &  0.746         \\

\hline
\hline
\end{tabular} \vspace{-3mm}
\end{table}

\subsection{Evaluation on Mass Classification}
We validate our model for two main mass classification tasks: PDAC vs. nonPDAC and patient management. SMCN is compared to the radiomics model and 2D/3D deep classification networks.  
We extract 2,410 radiomics features (482 for each CT phase) using Pyradiomics package~\cite{van2017computational} from the manually-annotated masses. These features include mass characteristics of various-ordered texture and shape descriptions. Gradient Boosting Decision Tree is used as the classifier~\cite{ke2017lightgbm}. Feature importance is calculated within each phase to select the most informative features (top 30) in the training fold. To compare with deep classification models, we use automatic mass segmentation and prepare 2D/3D tumor patches/volumes from all five phases. Each 2D patch is the tumor region representing the largest size in the axial view of the 3D volume which is then resized to $256 \times 256$ following typical texture analysis practices~\cite{Zhang_2017_CVPR}. For 3D volumes, we test with ResNet3D~\cite{hara3dcnns}, while for 2D patches we built deep texture networks using ResNet-18~\cite{he2016deep} as the backbone, following the model ~\cite{xue2018deep}. 


{\bf PDAC vs. NonPDAC.} We first evaluate our method for classifying PDAC vs. nonPDAC, which is the primary clinical diagnosis task. Comparative results are provided in Table \ref{tab:acc2}. Sensitivity and specificity correspond to the proportions of the correctly predicted PDACs and nonPDACs, respectively. We also conduct an ablation study on several classification strategies. Pixel voting (PV) indicates that the classification result is voted by the pixels from the segmentation masks -- using an optimal volume threshold from the validation set; vertex-based voting (VV) means that the result is voted by the classified vertices of Graph-ResNet; Global classification (GC) is our final model. All of our SMCN variations significantly outperform both 2D/3D deep networks, thus revealing that using only the tumor's texture information to distinguish PDAC from nonPDAC is insufficient. Our fully automated models greatly outperform the radiomics approach (sensitivity 0.853, specificity 0.868) using manual annotations. The SMCN-w-GC configuration reports the best quantitative results with a sensitivity of 0.945 and specificity of 0.906.


\begin{figure}
    \centering \vspace{-2mm}
    \includegraphics[width=0.98\linewidth]{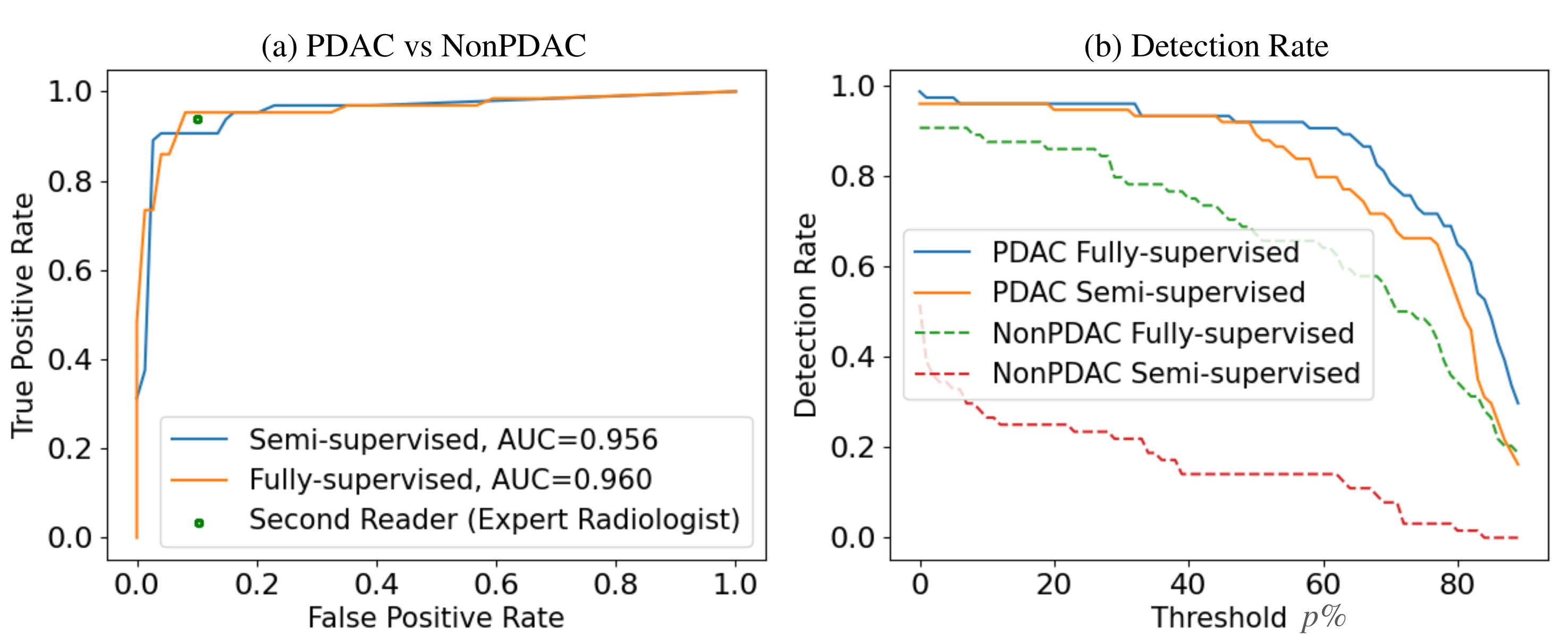}
    \caption{\small Comparison of (a) classification and (b) detection performances between fully-supervised and semi-supervised learning. The green dot depicts the performance of the mean second reader.}
    \label{fig:roc} \vspace{-5mm}
\end{figure}

{\bf Fully-supervised or Semi-supervised Learning.} As described in Sec.~\ref{sec:seg}, we can perform the PDAC vs. nonPDAC task even when only the patient-level nonPDAC labels are available (referred as semi-supervised). A fully-supervised setting means that both PDAC and nonPDAC annotated masks are obtained. ROC curves of PDAC (positive class) vs. NonPDAC (negative class) are shown in Fig.\ref{fig:roc}(A). Generally, similar classification results are achieved by the fully-supervised or semi-supervised learning method. 
Using only pseudo nonPDAC masks in a semi-supervised setting slightly reduces the detection rate of PDAC. Our fully-supervised method obtains sensitivity and specificity comparable to the mean second reader at a high-volume pancreatic cancer institution at 94\% and 90\%, respectively. 
The detection rates by different thresholds are evaluated in Fig.\ref{fig:roc}(b). A segmentation result is counted as successful detection if at least the fraction $p$ of the mass is segmented ($p$ is the cut-off). For fully-supervised learning, the best PDAC detection rate is 0.986 ($p>0$); the best nonPDAC sensitivity is 0.906 ($p \simeq 0.08$). Under semi-supervised learning, these numbers are 0.959 ($p\simeq 0.2$) and 0.515 ($p>0$), revealing the limitation of semi-supervised setting.

{\bf Quantitative Patient Management.} Patient management decisions fall under three categories: ``Surgery'', ``Monitoring''and ``Discharge'' (Fig.\ref{fig:ont}). We once again compare our methods to the Radiomics based approach, which is the current clinical standard, and conduct the ablation study on different classification strategies: PV, VV, and GC (Table \ref{tab:acc3}). Our SMCN method generally outperforms the radiomics method, and the SMCN-w-GC version yields the best results. Importantly, CompCyst \cite{springer2019multimodality} reports real pancreatic mass patient management results leveraging  clinical features, imaging characteristics, cyst fluid genetic and biochemical markers. However, it is invasive, and more expensive and time-consuming than SMCN. For a relatively fair comparison, we adhere to the same pancreatic disease taxonomy as \cite{springer2019multimodality}, i.e., exclude AC, DC, and RARE classes. Two confusion matrices, by CompCyst \cite{springer2019multimodality} and SMCN, are compared side-by-side in Table \ref{tab:cmfu}. 
SMCN achieves similar or slightly improved quantitative performance via  multi-phase CT imaging. (1) 95\% of {\bf surgery} patients (increased from 91\% by CompCyst) are correctly guided to surgery, but 1\% of surgery patients are misclassified as discharge patients, which is clinically undesirable. 
(2) The correctly recommended {\bf monitoring} patients have increased to 63\% by SMCN from 49\% in \cite{springer2019multimodality}. The error rate for ``recommending surgery'' has decreased significantly from 51\% \cite{springer2019multimodality} to 26\%. (3) The performance for {\bf discharge} patients are similar between SMCN and  CompCyst clinical results. We classify more patients into the surgery class (17\% vs. 13\%). 
This is due to the fact that the distribution of our patients among all three actionable classes is more unbalanced than that in \cite{springer2019multimodality}, and our {\bf discharge} patient sub-population may be insufficient. Finally, we provide patient-level classification results for all ten pancreatic tumor classes in the supplementary material (considered as clinically less critical). 




\vspace{-2mm}
\section{Conclusion} 

In this paper, we present a new segmentation-mesh-classification deep network (SMCN) to tackle the challenging and clinically-demanding tasks of pancreatic mass segmentation, diagnosis, and disease management. SMCN is composed of an anatomy-mass segmentation network, a mask-to-mesh 3D geometry modeling network, and a Graph-ResNet with vertex feature pooling and can be trained end-to-end. We extensively evaluate our SMCN method on the largest pancreatic multi-phase CT dataset of 661 patients, using the full taxonomy of ten types of pancreatic masses.
Our experiments demonstrate the superiority of the proposed SMCN for all three pancreatic analysis tasks.

{\small
\bibliographystyle{ieee_fullname}
\bibliography{egbib}
}

\newpage
\begin{center}
\textbf{\Large Supplemental Materials}
\end{center}

\section*{Full-Spectrum Taxonomy of Pancreatic Masses With Multi-Phase CTs, Labels, and Automated Results}

\begin{figure*}
\centering
\includegraphics[width=\linewidth]{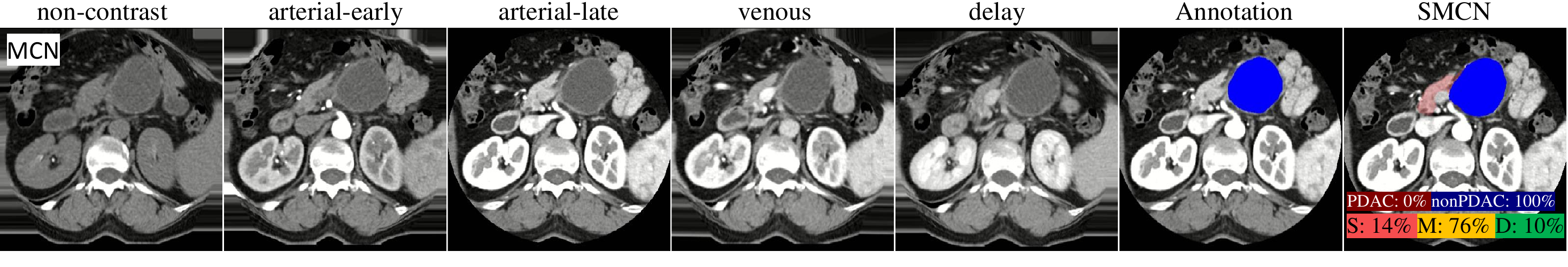}\\\vspace{-3mm}
\includegraphics[width=\linewidth]{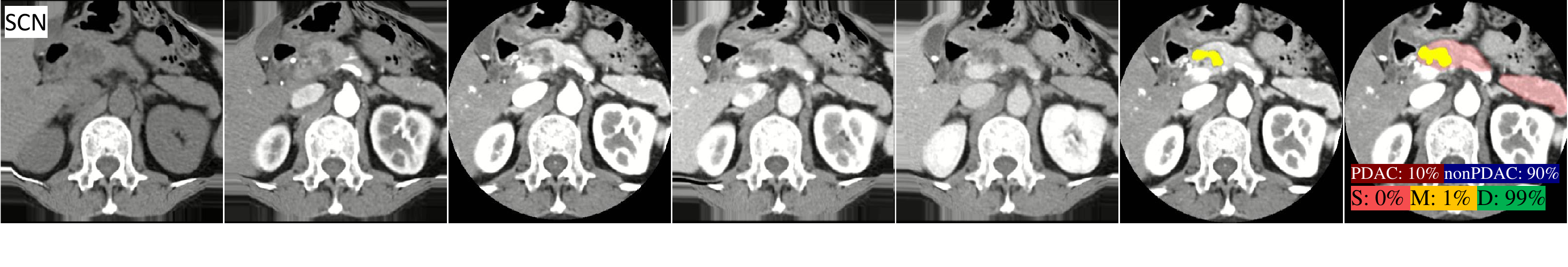} 
\caption{Visualized examples of full ten-class pancreas-mass qualitative segmentation results. Light red: pancreas. Green: ``Surgery'' mass (PDAC, AC, DC, PNET RARE, SPT, IPMN-high). Blue: ``Monitoring'' mass (IPMN, CP, MCN). Yellow: ``Discharge'' mass (SCN). All five CT imaging phases are shown as non-contrast, arterial-early, arterial-late, venous and delay. The annotation mask labels and segmentation results are overlaid on the arterial-late phase. The PDAC vs. NonPDAC classification probability and the patient management classification probability, ``Surgery'', ``Monitoring'' and ``Discharge'' are represented as S, M and D, are reported, respectively.}
\label{fig:5phase2} 
\end{figure*}

The inputs of  our SMCN model are 5-phase 3D CT scans for each patient:  non-contrast, arterial-early, arterial-late, venous, and delay phases, ordered by the acquisition time during the CT imaging process. The five phases are shown in Fig.\ref{fig:5phase2}. Firstly, the non-contrast phase is generated with less contrast. Then the Intravenous-injected contrast media is gradually applied or supplied into human body to increase the visibility of blood vessels (i.e., hyper-intensity patterns).  In the arterial-early phase, the arterial vessels become bright. In the arterial-late phase, the pancreas has a properly high imaging contrast which makes it the most suitable phase to observe the mass and perform diagnosis. In the venous phase, the venous vessels become bright in turn. Finally, in the delay phase, the brightness in all blood vessels vanishes and the kidneys become bright. Only the arterial-late phase has a high resolution on the transverse plane (median spacing is $0.419\times 0.419 \times 3$mm in [X, Y, Z]). The other four phases has a relatively lower resolution on the transverse plane (median spacing is $0.682\times 0.682 \times 3$mm). CT scans from the other four phases are spatially registered to the arterial-late phase. The input image size for inference is at the scale of $5 \times 512\times 512 \times 54$ voxels.

Fig.\ref{fig:5phase2} illustrates the automated results for qualitative evaluation. SMCN can guide the pancreas-mass segmentation network by integrating geometry cues and achieve more complete segmentation results, especially for harder and relatively rare nonPDAC classes. Our SMCN model can detect both small masses as in the DC and PNET cases, and large masses as in the SPT and MCN cases. In AC case, our SMCN model can detect the tumor which is outside of the pancreas head. In PNET and IPMN scenarios, our SMCN model detects (by segmentation) both the mass and the dilated duct (depicted as blue). In the IPMN-high case, our SMCN model labels both the low-grade IPMN mass (blue) and the high-grade IPMN mass (green) successfully. All examples collectively demonstrate that our graph-based anatomy geometry-aware SMCN network produces superior segmentation results.

\section*{The Spatial Distributions of Anatomic Locations of Ten-class Pancreatic Masses}

\begin{figure*}
\centering
\includegraphics[width=0.8\linewidth]{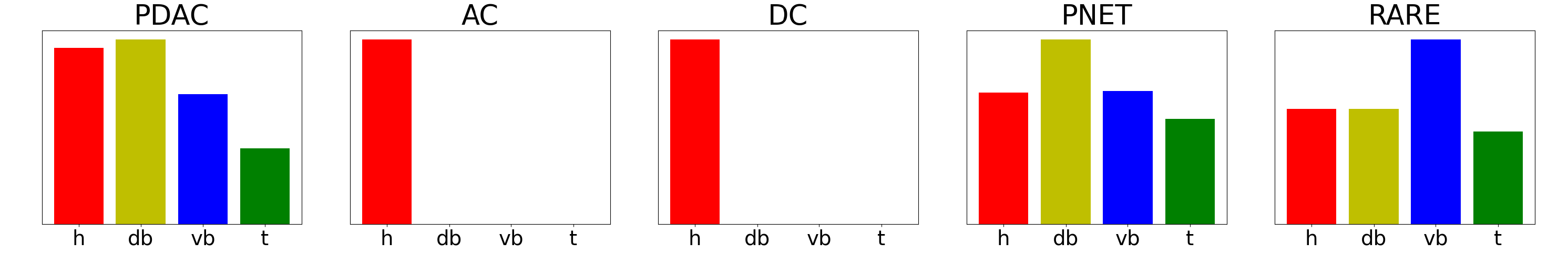}\\
\includegraphics[width=0.8\linewidth]{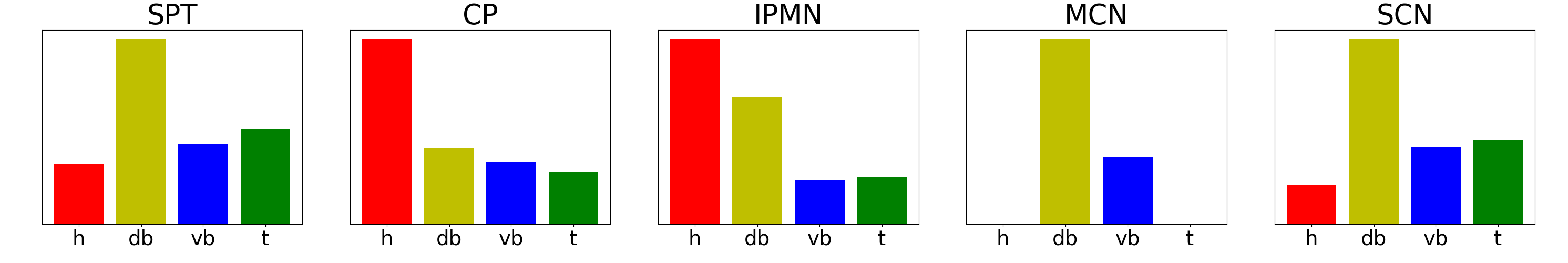} \vspace{2mm}
\caption{The spatial location distributions  of different (10) types of tumors associated with four pancreas parts. Red: pancreas head (h); Yellow: pancreas dorsal body (db); Blue: pancreas ventral body (vb); Green: pancreas tail (t).}
\label{fig:dis4} 
\end{figure*}

Our 3D mask-to-mesh algorithm can automatically divide the segmented pancreas into four anatomical regions or sub-regions of the pancreas: head, ventral body, dorsal body and tail, which are color-coded as red, blue, yellow and green in Fig. \ref{fig:dis4}, respectively.
 AC and DC only appear at the head; MCN does not appear on the head; CP and IPMN appear mostly on the head; PDAC, PNET and RARE are distributed over the entire surface of the pancreas. Cystic masses (SPT, SCN and MCN) are distributed over the entire pancreas with the majority being located at the dorsal body. These observed distributions verify the prior knowledge of different masses' spatial locations and motivate our geometry-aware mass diagnosis framework.

\section*{Full Ten-class Pancreatic Mass Quantitative Classification Results}
The patient-level 10-class pancreatic mass classification results are provided in Table \ref{tab:10}. Dominant tumor types achieve good accuracies (e.g., PDAC, IPMN, SPT), but others have relatively low performance, especially for DC and RARE. The imaging appearance of DC is very confusing against PDAC and occurs statistically rarely. RARE labeled patients have multiple co-existing diseases, which makes the diagnosis more complexly daunting. Precisely classifying the ten pancreatic tumor classes is a notoriously difficult long-tail disease recognition and reasoning problem (but is also clinically less important or critical). We would like to  explore this problem in the future work. 


\begin{figure*}
\centering
\includegraphics[width=\linewidth]{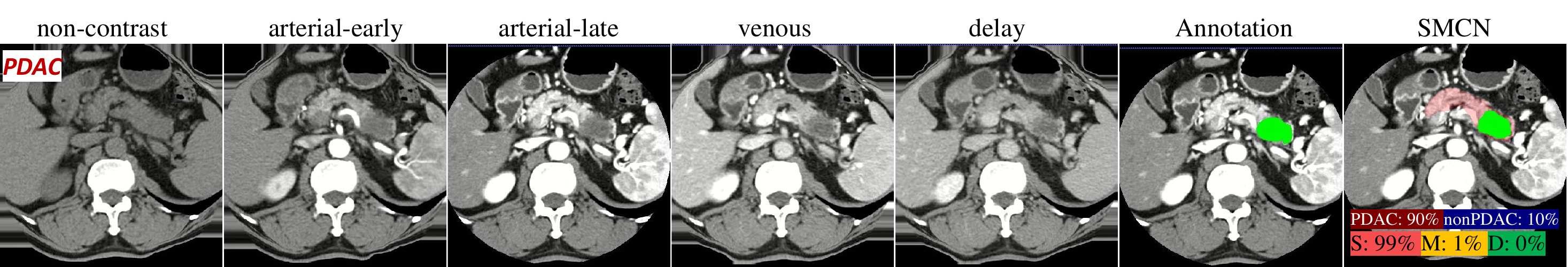}\\
\includegraphics[width=\linewidth]{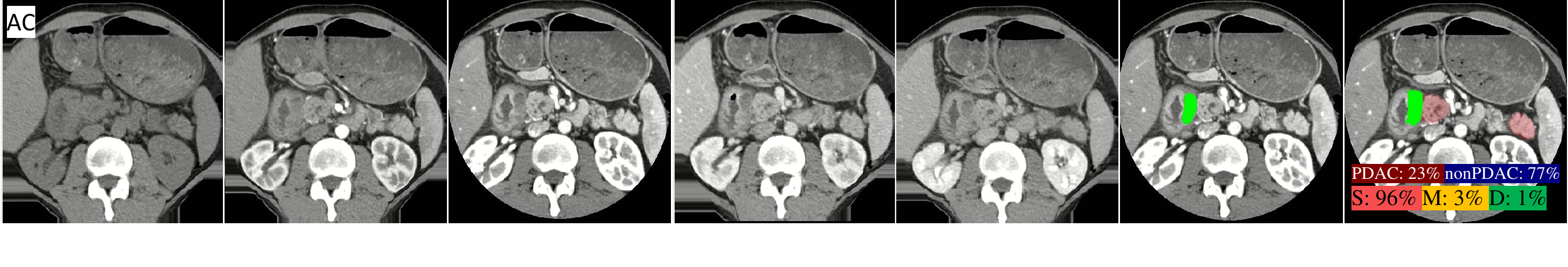}\\\vspace{-5mm}
\includegraphics[width=\linewidth]{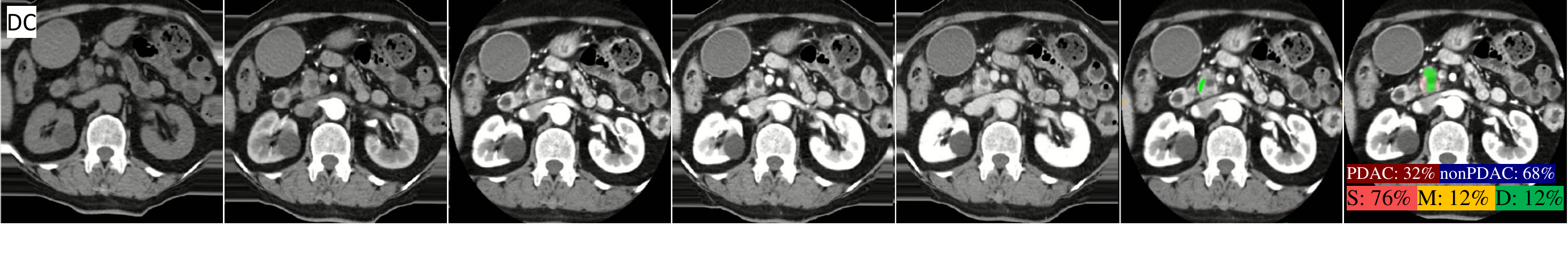}\\\vspace{-5mm}
\includegraphics[width=\linewidth]{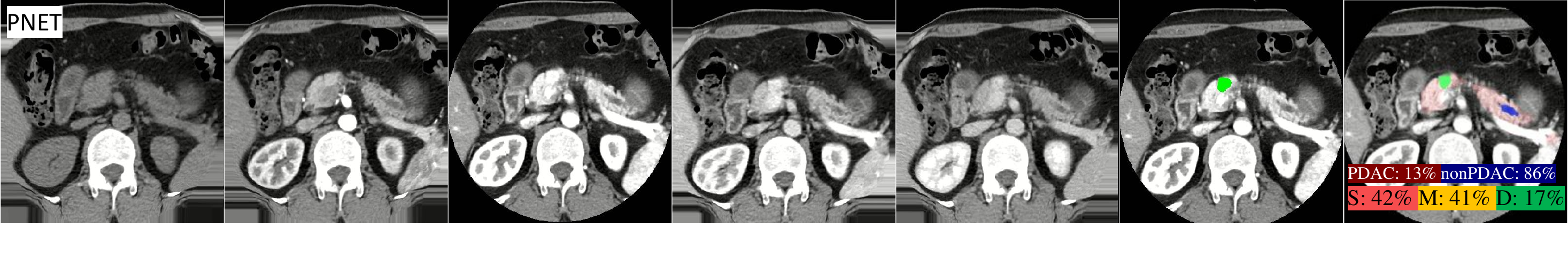}\\\vspace{-5mm}
\includegraphics[width=\linewidth]{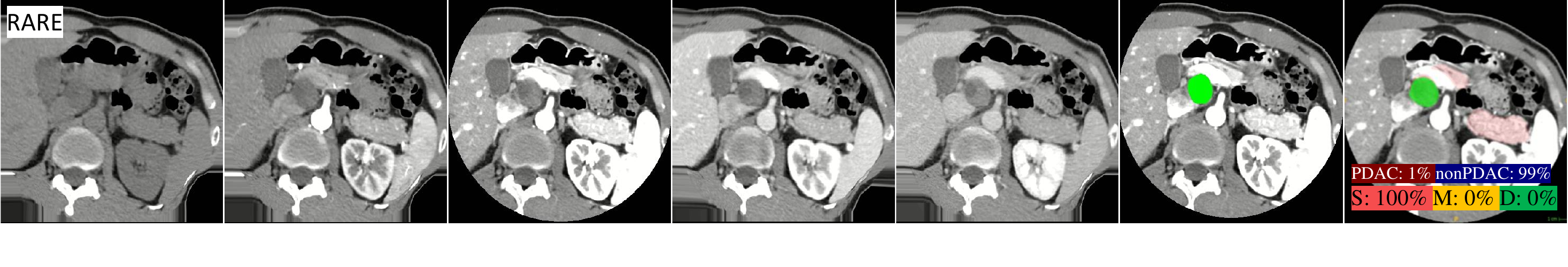}\\\vspace{-5mm}
\includegraphics[width=\linewidth]{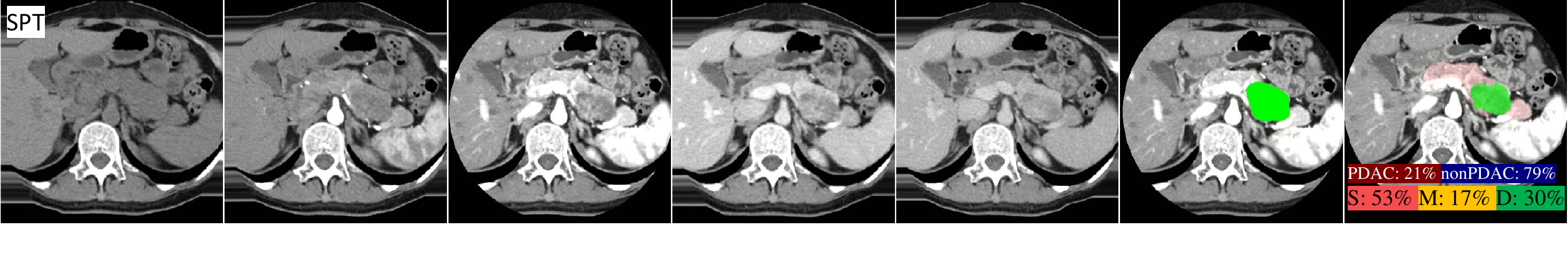}\\\vspace{-5mm}
\includegraphics[width=\linewidth]{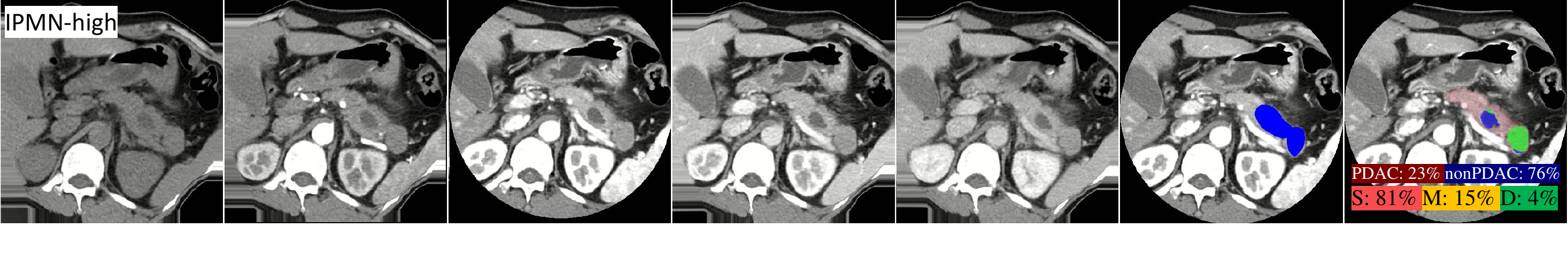}\\\vspace{-5mm}
\includegraphics[width=\linewidth]{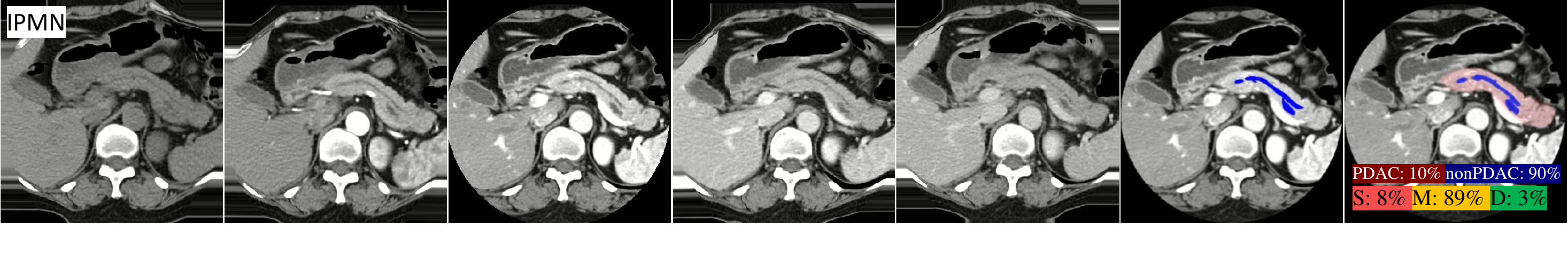}\\\vspace{-5mm}
\includegraphics[width=\linewidth]{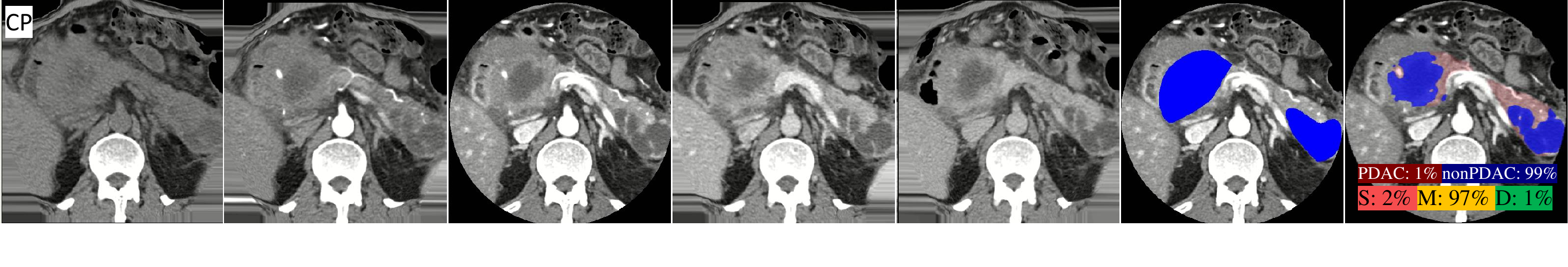}\vspace{-5mm}
\label{fig:5phase} \vspace{-3mm}
\end{figure*} 

\begin{table*}
\centering

\begin{tabular}{ c  | c|  c|c|c|c|c|c|c|c|c|c }
\hline 
class     & PDAC &AC & CP& DC & IPMN & MCN  & PNET & RARE & SCN & SPT & Avg. \\
\hline

recall &0.949& 0.488&  0.473& 0 & 0.759 & 0.333& 0.330 & 0& 0.533 &  0.420  & 0.428 \\
\hline
precision &0.904 & 0.623& 0.663&      -& 0.427 &0.167 & 0.426&    -& 0.515& 0.733 & 0.557\\
\hline
\end{tabular} \vspace{2mm}
\caption{Ten-class patient-level pancreatic disease (mass) classification results. Dominant tumor types achieve good accuracies (e.g., PDAC, IPMN, SPT), but others have relatively low performance, especially for DC and RARE. The imaging appearance of DC is very confusing against PDAC and distributes very rarely. RARE patients have multiple co-existing diseases, which makes the diagnosis process more complexly daunting or confusing. Classifying all ten pancreatic tumor/mass classes with high precision is a notoriously difficult, long-tail recognition and reasoning problem. Fortunately, this task is also clinically less important.
}
\label{tab:10}
\end{table*}

\end{document}